\documentclass[12pt,preprint]{aastex}
\def\be{\begin{equation}}
\def\ee{\end{equation}}

\def\msun{M_{\odot}}

\def\be{\begin{equation}}
\def\ee{\end{equation}}
\catcode`\@=11 % This allows us to modify PLAIN macros.
\def\@versim#1#2{\vcenter{\offinterlineskip
        \ialign{$\m@th#1\hfil##\hfil$\crcr#2\crcr\sim\crcr } }}
\def\lsim{\mathrel{\mathpalette\@versim<}}
\def\gsim{\mathrel{\mathpalette\@versim>}}
\def\mpy{M_\odot \ {\rm yr^{-1}}}

\shorttitle{Testing RIAF model for Sgr A*}
\shortauthors{Yuan et al.}
\begin{document}
%\date{}

\title{Testing RIAF model for Sgr A* using the size measurements}

\author{Feng Yuan\altaffilmark{1,2}, Zhi-Qiang Shen\altaffilmark{1,2} 
and Lei Huang\altaffilmark{1,3}}
\altaffiltext{1}{Shanghai Astronomical Observatory, Chinese Academy of
Sciences, 80 Nandan Road, Shanghai 200030, China;fyuan,zshen,muduri@shao.ac.cn}
\altaffiltext{2}{Joint Institute for Galaxy and Cosmology (JOINGC) of SHAO 
and USTC}
\altaffiltext{3}{Graduate School of Chinese Academy of Sciences, Beijing 100039,
China} 

\begin{abstract}

Recent radio observations by the VLBA at 7 and 3.5 mm produced the
high-resolution images of the compact radio source located at the
center of our Galaxy---Sgr A*, and detected its
wavelength-dependent intrinsic sizes at the two wavelengths. This
provides us with a good chance of testing previously-proposed
theoretical models for Sgr A*. In this {\em Letter},
we calculate the size based on the radiatively inefficient
accretion flow (RIAF) model proposed by Yuan, Quataert \& Narayan
(2003). We find that the predicted sizes after taking into account
the scattering of the interstellar electrons are
consistent with the observations. We further predict an image of
Sgr A* at 1.3 mm which can be tested by future observations.

\end{abstract}


\section{Introduction}

The compact radio source located at the center of our Galaxy, Sgr A*,
is perhaps the most intensively studied black hole source up to date (see review
by Melia \& Falcke 2001). Substantial observational results put strict
constraints on theoretical models. These models include the spherical
accretion model (Melia, Liu \& Coker 2001; Liu \&
Melia 2002), the pure jet model (Falcke et al. 1993; Falcke \& Markoff 2000),
the advection-dominated accretion flow (ADAF) or radiatively inefficient
accretion flow (RIAF) (Narayan et al. 1995;
Narayan et al. 1998; Yuan, Quataert \& Narayan 2003, 2004), and the
coupled ADAF-jet model (Yuan, Markoff \& Falcke 2002). In the present
paper we concentrate on the RIAF model proposed by Yuan,
Quataert \& Narayan (2003, hereafter YQN03).

The YQN03 model explains most of the observations available at
that time, including the spectrum from radio
to X-ray, the radio polarization, and the flares at both infrared
and X-ray wavebands (see YQN03 for detail).
After the publication of YQN03, many new observations are
conducted. These include new spectral variability at
millimeter wavelength (Zhao et al. 2003; Miyazaki et al. 2004; Mauerhan 
et al. 2005; An et al. 2005), the high angular resolution
measurements of the linear polarization at submillimeter wavelengths 
and its variability with SMA (Marrone et al. 2005), and very high energy 
emissions from the direction of Sgr A* (INTEGRAL: B\'elanger et al.
2004; HESS: Aharonian et al. 2004; CANGAROO: Tsuchiya et al.
2004; MAGIC: Albert et al. 2006). Several large multiwavelength 
campaigns have been performed (e.g., Eckart et al. 2004, 2005; Yusef-Zadeh
et al. 2005). Some observations mentioned above confirm
the YQN03 model, or they can be easily interpreted
in the context of this model, while some observational results are
not so obvious to be understood and thus offer new challenges to
the model. In the present {\em Letter} we will discuss the size of Sgr A* 
at radio wavelengths, which has not been discussed in YQN03.

It has long been realized that due to the effect of scattering by
the interstellar electrons, the intrinsic size of Sgr A* is only
detectable at short wavelength (Davis, Walsh \& Booth 1976; Lo et
al. 1985, 1998; Krichbaum et al. 1997; Bower \& Backer
1998). This is because the scattering theory shows that at long
wavelength the observed image size will be dominated by the
scattering and scale quadratically as a function of wavelength
(Narayan \& Goodman 1989). At short wavelength, however, precise
measurements of the size of Sgr A* are seriously hampered by
calibration uncertainties. Recently, great progress has been made
in this aspect due to the improvement of the model fitting
procedure by means of the closure amplitude. Using the VLBA, at 7
mm wavelength, Bower et al. (2004) successfully measured the size
of Sgr A* of $0.712^{+0.004}_{-0.003}$ mas, Shen et al. (2005) 
obtained averaged size of 0.724 $\pm 0.001$ mas and 
$0.21^{+0.02}_{-0.01}$ mas at 7 and
3.5 mm, respectively. By subtracting in quadrature the scattering
size, they obtained an intrinsic size of $0.237\pm 0.02$ mas
(Bower et al. 2004) or 0.268 $\pm0.025$ mas (Shen et al. 2005) at
7 mm and 0.126 $\pm 0.017$ mas at 3.5 mm (Shen et al. 2005).
Since this new constraint is independent of the other observations 
such as spectrum and variability, it provides an independent test to 
investigate whether or not the RIAF model proposed by YQN03 can account 
for the observed sizes.

\section{RIAF Model for Sgr A*}

We first briefly review the RIAF model of YQN03, which can be
considered as an updated version of the original ADAF model for
Sgr A* (Narayan et al. 1995; 1998). Compared to the ADAF model,
two main developments in the RIAF model are the inclusions of
outflow/convection and the possible existence of nonthermal
electrons. The former is based on the theoretical calculations and
numerical simulations (e.g., Stone et al. 1999; Hawley \& Balbus
2002). The possible existence of
nonthermal electrons is due to the acceleration processes such as
turbulent acceleration, reconnection, and weak shocks in accretion
flow. We characterize the nonthermal population by $p$ [$n(\gamma) \propto
\gamma^{-p}$ where $\gamma$ is the Lorentz factor], and a
parameter $\eta$, the ratio of the energy in the power-law
electrons to that in the thermal electrons. The dynamical
quantities describing the accreting plasma, such as the density
and temperature, are obtained by globally solving a set of
accretion equations including the conservations of fluxes of mass,
momentum, and energy. We assume that the accretion rate is a
function of radius, i.e., $\dot{M}=\dot{M}_0 (R/R_{\rm out})^s$
(e.g., Blandford \& Begelman 1999).  Here $R_{\rm out}$ is the
outer radius of the flow, i.e., the Bondi radius, $\dot{M}_0$ is
the accretion rate at $R_{\rm out}$ (the Bondi accretion rate,
fixed by {\em Chandra} observations of diffuse gas on $\sim 1''$
scales; Baganoff et al. 2003). The radiative processes
we considered include synchrotron, bremsstrahlung and their
Comptonization by both thermal and nonthermal electrons. The sum
of the self-absorbed synchrotron radiation from the thermal
electrons at different radii dominates the radio emission of Sgr
A* at $\ga 86$ GHz, while the radio emission at $\la 86$ GHz is
the sum of the synchrotron emission of both thermal and nonthermal
electrons. As we stated in YQN03, there is no much freedom in the 
choice of parameter values in the RIAF model. 

To calculate the intrinsic size of Sgr A* predicted by the RIAF
model and compare with observations, we need to adjust the mass of
the black hole. The mass of the black hole adopted in YQN03 is
$2.5 \times 10^6 \msun$. Recent observations show that the mass
should be larger---$M/\msun=3.7\pm 1.5, 3.3\pm0.6$, and $4.1\pm0.6
\times 10^6$ in Sch\"odel et al. (2002, 2003), and Ghez et al.
(2003), respectively. We adopt $M=4 \times 10^6\msun$. 
Thus the model parameters need to be adjusted accordingly to
ensure that the adjusted model can fit the spectrum of Sgr A*
equally well. The new parameters are: $\dot{M}_0\approx 10^{-6}
\mpy, s=0.25$, $\eta=0.4\%$, and the fraction of the turbulent energy
directly heating electrons $\delta=0.3$.
We note that the values of $\dot{M}_0$, $s$ and $\eta$ change
little, but the value of $\delta$ decreases from 0.55 in YQN03 to
the present 0.3. This is because the electron temperature needs to
decrease a bit to compensate for the increase of flux due to the
increase of the mass of the black hole.

\section{The size of Sgr A* predicted by the RIAF model}

The observed radio morphology of Sgr A* is broadened by the
interstellar scattering, which is an elliptical Gaussian along a
position angle of $\sim 80\arcdeg$ with the major and minor axis
sizes in mas of $\theta_{\rm scat}^{\rm maj}=(1.39\pm0.02)
\lambda^2$ and $\theta_{\rm scat}^{\rm
min}=(0.69\pm0.06)\lambda^2$, respectively (Shen et al. 2005). The
observing wavelength $\lambda$ is in cm. To get the intrinsic size
of Sgr A*, observers have to subtract the scattering effect from
the observed image. Here, all the sizes estimated from
observations are referred to as the FWHM (Full Width at Half
Maximum) of the Gaussian profile. This requires that not only the
observed apparent image, but the intrinsic intensity profile of
the source can be well characterized by a Gaussian distribution.
However, this may not necessarily be the case. For Sgr A*, we will
show that the intrinsic intensity profile emitted by the RIAF can
be quite different from the Gaussian distribution. In this case,
we are unclear to the definition of the ``intrinsic size'', let
alone the comparison between the theoretically predicted size and
the observationally derived one.
%This casts doubt on previous simple calculations on the
%size of Sgr A* (e.g., Ozel et al. 2000).
Given this situation, in the present paper we will not try to
calculate the ``intrinsic'' size of Sgr A*. Rather, we first
calculate the intrinsic intensity profile from the RIAF model.
Then we take into account the scatter broadening toward the
Galactic center to obtain the simulated image. We will directly
compare the simulated image with the observed one.

Now let's calculate the specific intensity profile of the
radiation from the RIAF. We first assume that the black hole in
Sgr A* is non-rotating and the RIAF is face-on. The effects of the
assumptions on the result will be discussed later. We first solve
the global solution to obtain the dynamical quantities of the RIAF
as stated in Section 2. Because in our calculation the Paczy\'nski
\& Wiita (1980) potential is used and the calculation is in the
frame of Newtonian mechanics rather than the exact general
relativity (GR), the calculated radial velocity of the accretion
flow very close to the black hole is larger than the speed of
light thus not physical. As a result, at this region the density
of the accretion flow is smaller and correspondingly the electron
temperature is also lower due to weaker compression work. To
correct this effect, for simplicity we compare the radial velocity
obtained in our calculation with that obtained by Popham \& Gammie
(1998) in the frame of GR. We found that our radial velocity at
$r\la 30$ should be divided by $0.93 e^{2.13/r}$ where $r$ is the
radius in unit of $R_g(\equiv GM/c^2)$. As for the electron
temperature, following the result in Narayan et al. (1998), a
correction factor of $1.4r^{0.097}$ is adopted. The above
corrections are of course not precise, but fortunately the result
is not sensitive to them as we will discuss in Section 4.

The resulting intrinsic intensity profiles at 3.5 and 7 mm are
shown by the red solid lines in Fig. 1(b)\&(f). Obviously, these
two profiles can't be well represented by a Gaussian distribution.
Before we incorporate the electron scattering, however, we take
into account the following additional relativistic effects, namely
gravitational redshift, light bending, and Doppler boosting
(Jaroszynski \& Kurpiewski 1997; Falcke et al. 2000). We implement
these effects using our GR ray-tracing code (Huang et al. in
preparation). The dashed lines in Fig. 1(b)\&(f) show the
resultant intensity profiles after the above GR effects are
considered. The original peak of each solid line becomes lower
because of the strong gravitational redshift near the black
hole. The outward movement of the peak location is due to light
bending.

Fig. 1(c)\&(g) show the simulated image after the scattering has
been included. The scattering model mentioned at the beginning of
this section is adopted. The images are elliptical, consistent 
with observations. The open circles in Fig. 1(d)\&(h) show
the intensity of the simulated image as a function of radius. The
smoothness of the profile is because of the scattering broadening.
The solid lines in Fig. 1(d)\&(h) are Gaussian fit to the open
circles. It can be seen that the intensity profile of the
simulated image can be perfectly fitted by a Gaussian, as we
stated above. The FWHM of the simulated images at 7 mm and 3.5 mm
are $0.729^{+0.01}_{-0.009}$ mas and $0.248^{+0.001}_{-0.002}$
mas, respectively. The simulated size at 7 mm is in good agreement
with the observed value by Shen et al. (2005) within the error
bars but slightly larger than the observed size by Bower et al.
(2004); the size at 3.5 mm is a little larger than the observation
of Shen et al. (2005). Given that the size of the source may be
variable (Bower et al. 2004) and the uncertainties in our
calculations that we will discuss in \S4, we conclude that the
predictions of the YQN03 model are in reasonable agreement with
the size measurements. 

In the above simulation, the ``input'' intensity profile for the
scattering simulation is the result of considering various effects
or corrections. In the following we discuss the effects of these
corrections by considering various ``input'' intensity profiles.
The first profile we consider is the one without the GR effect,
i.e., the red solid lines in Fig. 1(b)\&(f). In this case, the
FWHM of the simulated image after considering electron scattering
are 0.737 and 0.239 mas at 7 and 3.5 mm, respectively. So the GR
effects make the size of Sgr A* slightly larger at 3.5 mm. This is
because the strong GR effects make the emission very close to the
black hole weaker, while the emission at large radii almost remain
unchanged. But at 7 mm, since the scattering effect is much
stronger (4 times) than at 3.5 mm, the emission at both the small
and large radii in the scattered intensity profile becomes
weaker due to the GR effects. The total effect is that the size becomes
smaller at 7 mm. We have confirmed our interpretation by simulating the
image at a longer wavelength---14 mm. The second profile we consider is
based on the last profiles (i.e., without considering GR effects), with
the only difference that we now only consider the emission of thermal
electrons in calculating the intrinsic intensity profiles. The
FWHM values of the simulated image in this case are 0.724 and
0.228 mas at 7 mm and 3.5 mm, respectively. So the inclusion of
the nonthermal electrons in the RIAF makes the size of Sgr A* at 7
mm and 3.5 mm larger. This is because the intensity profile from
the nonthermal electrons are flatter than that of the thermal
electrons. The last input intensity profile we consider is based
on the second profiles above (i.e., without considering nonthermal
electrons) but with the difference that the profiles of the
density and electron temperature are directly obtained from the
global solution of RIAF and no relativistic corrections to the
profiles of density and temperature are adopted. In this case, the
FWHM values of the simulated image are 0.727 mas and 0.238 mas at
7 mm and 3.5 mm, respectively. So the inclusion of relativistic
corrections to the profiles of density and electron temperature
makes the size of Sgr A* smaller. This is because the corrections
make the emission at the innermost region of RIAF stronger.

We also calculated the simulated size of Sgr A* at 1.35 cm, which
is $2.67_{-0.03}^{+0.04}$ mas. This result is consistent with the 
observed size of $2.635^{+0.037}_{-0.024}$ mas by Bower et al. (2004) 
and slightly larger than the size of $2.53^{+0.06}_{-0.05}$ mas by Shen 
et al. (2005). At last we try to predict an observed image of Sgr A* 
at a shorter wavelength---1.3 mm. The red solid line in Fig.
1(j) shows the calculated intensity profile while the dashed line
is the profile after the GR effects are taken into account using
the ray-tracing method. The simulated image at 1.3 mm after
considering the electron scattering is shown in Fig. 1(k) and its
intensity profile is shown in Fig. 1(l). Different from the cases
of 7 mm (Fig. 1(d)) and 3.5 mm (Fig. 1(h)), however, the simulated
intensity profile can no longer be reasonably fitted by a Gaussian
(see also Fig. 1 in Falcke, Melia \& Agol 2000). This indicates
that the ratio between the intrinsic size and the scattering size
must be larger at 1.3 mm than that at 3.5 mm where the two sizes
are comparable. And as a result, the non-Gaussian distribution of the
intrinsic intensity distribution significantly modulate the
observed image. This prediction can be tested by future VLBI
observations.

\section{Summary and Discussion}

The recent VLBA observations (Shen et al. 2005; Bower et al. 2004)
produced high-resolution images of Sgr A* at wavelengths of 3.5
and 7 mm. The measured sizes provide a good chance of
testing theoretical models. In this Letter we investigate whether
the RIAF model presented in YQN03 can account for these new
observations. We calculate the intrinsic intensity profile of
RIAF, taking into account the relativistic corrections such as
light bending and gravitational redshift. Because the intrinsic
intensity profile produced by the YQN03 model can't be represented
by a Gaussian (ref. the solid lines in Fig. 1(b)\&(f)), we
simulate the image by considering the interstellar scattering. The
results are shown in Fig. 1(d)\&(h). The intensity profile of such
an image can be fitted by a Gaussian and we thus obtain its FWHM
value and compare it directly with the observations (Fig.
1(d)\&(h)). The predicted sizes of Sgr A* by RIAF model of YQN03
at 7 and 3.5 mm are $0.729^{+0.01}_{-0.009}$ mas and
$0.248^{+0.001}_{-0.002}$ mas respectively, which are in
reasonable agreement with observations considering the uncertainties
of the calculations. We further predict an image of Sgr A* at 1.3
mm (Fig. 1 (k)\&(l)) which can be tested by future observations.

In our calculations, we assume a face-on RIAF and a non-rotating
black hole. If the RIAF is not face-on, the result will be more
complicated, depending on the angle between our line of sight and
the rotation axis of RIAF, and the angle between the major axis of
the scattering screen and the rotation axis of the RIAF.
Quantitative estimation of the size in this case is thus
difficult. But given the geometry of RIAF, we speculate that the
results should be similar, even at the extreme case of an edge-on
RIAF. If the black hole is rapidly rotating, however, the
accretion flow will extend inward further compared to the case of
a non-rotating Schwarzschild hole, thus the peak in the intensity
profile (ref. Fig. 1(b)\&(f)) will move to smaller radii and its
amplitude will become higher. This will result in a somewhat
smaller size of Sgr A*. Given that the predicted sizes at both 3.5
and 7 mm by the RIAF model around a Schwarzschild black hole are
larger than observations, our calculations thus suggest
that the black hole in Sgr A* may be rapidly rotating. The exact
prediction of the angular momentum of the black hole needs fully
self-consistent radiation-hydrodynamics calculations to both the
dynamics and the radiation of RIAF in the Kerr geometry, which is
beyond the scope of the present paper.

At last we briefly discuss the constraint of the observed size of
Sgr A* on other two models of Sgr A*, namely the jet model of
Falcke \& Markoff (2000) and the coupled jet-ADAF model of Yuan,
Markoff \& Falcke (2002). One main difference between these two
models associated with the present paper is that in the former the
radio emission above $\sim 86 $GHz is produced by the nozzle of
the jet while in the latter the contribution of the ADAF is
significant. Falcke \& Markoff (2000) calculated the size of Sgr
A*. The predicted sizes of Sgr A* at 3.5 mm by the nozzle and the
jet components are $\sim 0.04$ and $0.16$ mas, respectively. Since
in this model the emission at 3.5 mm is dominated by the nozzle
rather than the jet, the predicted size might be $\ga 0.04$ mas,
much smaller than the observed value. Numerical calculations are
required to confirm this speculation. On the other hand, in the 
jet-ADAF model, the contribution of the emission from the ADAF can 
dominate over that from the jet under suitable parameters. In this 
case, the predicted size of Sgr A* will be consistent with the
observations, as we show in the present paper. Of course, the ADAF
component there needs to be replaced by a RIAF, i.e., considering
the outflow/convection. In that case, the main difference between
the jet-ADAF model and the RIAF model is the origin of the radio
emission below $\sim$ 86 GHz.

\begin{acknowledgements}
This work was supported in part by the
One-Hundred-Talent Program and the National Natural Science
Foundation of China (grants 10543003 and 10573029).

\end{acknowledgements}

\clearpage

\begin{figure} \epsscale{1.0} \plotone{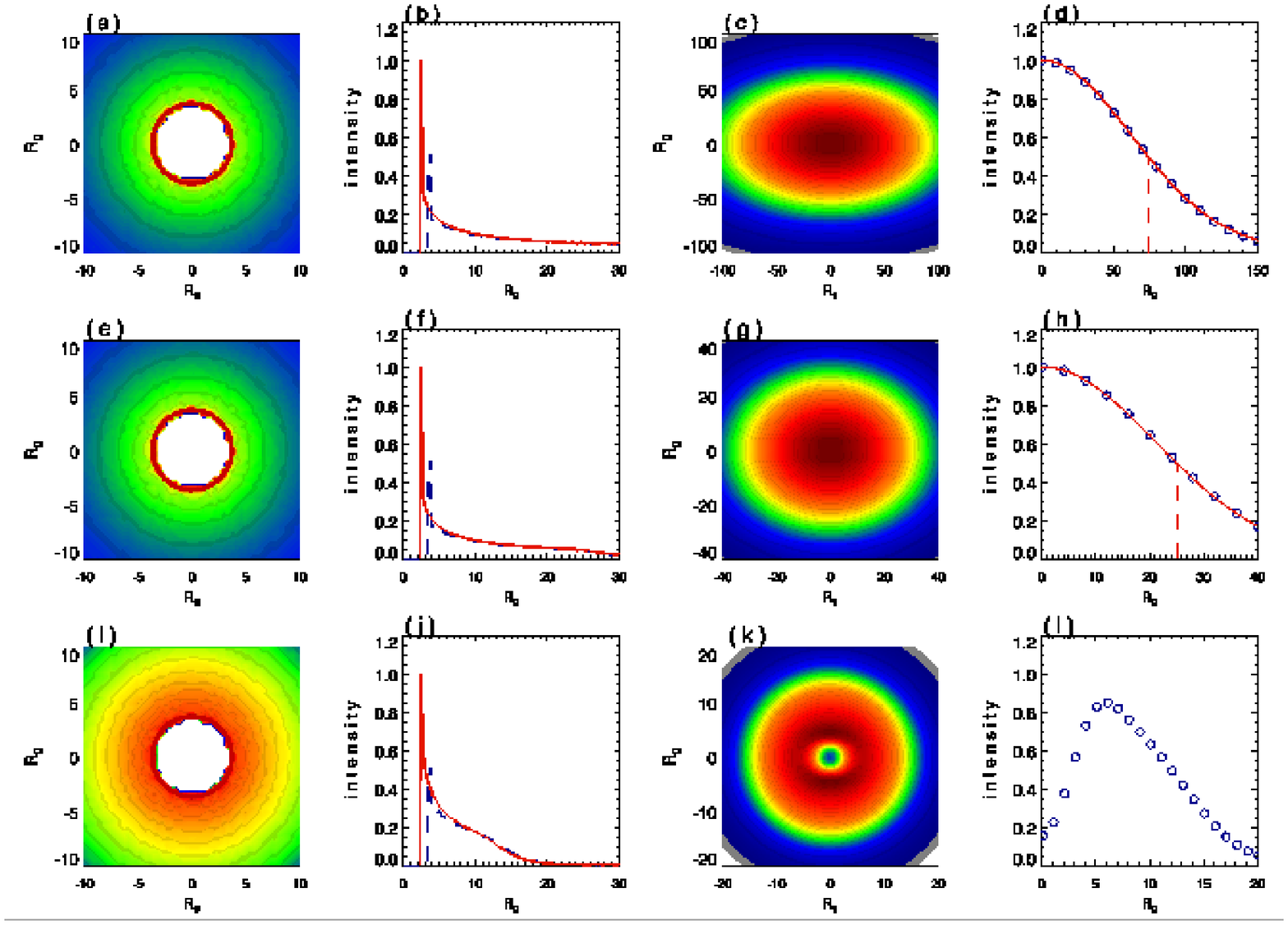} \vspace{.0in}
\caption{The images and sizes of Sgr A* at 7 (top panel), 3.5
(middle panel), and 1.3 mm (bottom panel). At each panel, the
first column shows the ``input'' intensity distribution. The solid
line in the second column is the intensity profile calculated from
the RIAF model while the dashed line shows the intensity profile
after the GR effects are taken into account using the ray-tracing
method. The third column shows the simulated image after the
interstellar scattering is taken into account. The open circles in
the fourth column shows the intensity profile of the simulated
image while the solid line shows the Gaussian fit to the circles.
The vertical dashed line shows the location of the FWHM. At 1.3
mm, the simulated profile can't be fit by a Gaussian, and thus no
FWHM is indicated. }
\end{figure}

\end{document}